# Characterizing Workload of Web Applications on Virtualized Servers

Xiajun Wang[1,2], Song Huang[2], Song Fu[2] and Krishna Kavi[2]
[1]Department of Information Engineering
Changzhou Institute of Light Industry Technology, China
[2]Department of Computer Science and Engineering
University of North Texas, Denton, Texas, USA
E-mail: dy_wxj@163.com, SongHuang@my.unt.edu, {Song.Fu, Krishna.Kavi}@unt.edu

## Abstract

With the ever increasing demands of cloud computing services, planning and management of cloud resources has become a more and more important issue which directed affects the resource utilization and SLA and customer satisfaction. But before any management strategy is made, a good understanding of applications' workload in virtualized environment is the basic fact and principle to the resource management methods. Unfortunately, little work has been focused on this area. Lack of raw data could be one reason; another reason is that people still use the traditional models or methods shared under non-virtualized environment. The study of applications' workload in virtualized environment should take on some of its peculiar features comparing to the non-virtualized environment. In this paper, we are open to analyze the workload demands that reflect applications' behavior and the impact of virtualization. The results are obtained from an experimental cloud testbed running web applications, specifically the RUBiS benchmark application. We profile the workload dynamics on both virtualized and non-virtualized environments and compare the findings. The experimental results are valuable for us to estimate the performance of applications on computer architectures, to predict SLA compliance or violation based on the projected application workload and to guide the decision making to support applications with the right hardware.

**Keywords**: Workload characterization, Virtualization, Performance modeling, Cloud computing.

## 1. Introduction

The increasingly popular cloud computing paradigm provides on-demand access to computing and storage with the appearance of unlimited resources [1]. Users are given access to a variety of data and software utilities to manage their work. Users rent virtual resources and pay for only what they use. Underlying these services are data centers that provide virtual machines (VMs) [2]. Virtual machines make it easy to host computation and applications for large numbers of distributed users by giving each the illusion of a dedicated computer system. It is anticipated that cloud platforms and services will increasingly play a critical role in academic, government and industry sectors, and will have widespread societal impact.

Resource planning and management is crucial for building cost-effective cloud systems and services with a high service-level agreement (SLA) and customer satisfaction rate. Current solutions to resource management usually over-provision VMs and/or their capacity to cloud applications [3]. However, a fundamental question, i.e., “What are the characteristics of applications’ runtime behavior on the cloud?” or “What impact does virtualization have on the resource demands from cloud applications?”, has not yet been answered. There exists research on analyzing the performance traces collected from data centers [4, 5]. Still, none of them evaluate the influence of virtualization on the applications’ resource demands in cloud computing infrastructures.

The goal of this work is to characterize runtime workload of cloud applications in the virtualized environment and compare it with traditional, non-virtualized systems. To the best of our knowledge, this is the first work to analyze the impact of virtualization on the resource demands of cloud applications. In this paper, we present the experimental results on a cloud testbed. We run an illustrating web application, i.e., RUBiS (Rice University Bidding System) benchmark [6], on cloud servers. We profile the application’s workload dynamics on both virtualized and non-virtualized environments. We compare the resource demands of CPU, RAM, disk and network at the three tiers (i.e., web, application and database servers) of RUBiS while serving thousands of client requests. The findings and knowledge will help us accurately estimate the performance of applications, predict SLA compliance or violation based on the projected application workload and guide the decision making to support applications with the right hardware in the cloud.

The rest of this paper is organized as follows. Section 2 discusses the related work. We describe the settings of the cloud testbed and the application benchmark in Section 3. The experimental results are presented in Section 4. Section 5 concludes the paper with remarks on the future work.

## 2. Related Work

Workload characterization studies are useful for helping system operators identify system bottlenecks and design solutions for performance optimization. Existing research efforts target different systems and components including data centers [4, 5], Web servers [7, 8], storage [9, 10, 11] and network [12, 13]. Several studies [14, 15, 16] focus on workload analysis in the grid and parallel computing systems. They present various methods for analyzing and modeling workload traces. However, the application characteristics and resource scheduling policies in high-performance computing (HPC) systems are different from those in the cloud [17, 18, 19].

Existing work on workload characterization can be classified into two major categories: model-driven and trace-driven methods. Model-driven approaches, such as [20], analyze resource utilization and application performance based on assumptions of workload distributions. The resource demand of a program is estimated by checking the types and number of instructions of the program and its structure. The overhead of modeling large and complex applications is prohibitive and the accuracy of the models is compromised by static analysis.

Trace-driven approaches study performance traces collected from real or controlled systems in order to discover the time series of user requests and resource usage. Distributions of profiled metrics are analyzed to describe workload characterization. For example, Kavulya et al. [21] analyze the job patterns and failure sources based on application execution traces from an HPC cluster. Mishra et al. [22] focus on the characteristics of resource demands on CPU and memory. The Yahoo Cloud Serving Benchmark [23] characterizes the activity of database-like systems at the read/write level. Their work focus on estimating application completion time and looking for performance problems based on application execution traces. Moreover, as applications display various workload dynamics, it is difficult to exploit this approach in capacity planning and real system analysis.

There is little work on understanding applications' workload dynamics in cloud computing environments. As virtualization has been an enabling technology for cloud computing, it is imperative to investigate the impact of virtualization on the resource demands of cloud applications, which is the focus of this work.

## 3. Cloud Testbed and Benchmark

The cloud computing system under test consists of HP ProLiant servers which are connected by gigabit Ethernet. Each cloud server is equipped with 8 Intel Xeon 2.8 GHz cores, 32 GB of RAM and 2 TB of disk. We have installed Xen 3.1.2 hypervisors on the cloud servers. The operating system on a virtual machine is Linux 2.6.18 as distributed with Xen 3.1.2. The cloud testbed is organized and built in an Amzon EC2-like [24] style providing IaaS cloud services. Each cloud server hosts up to ten VMs. A VM is assigned up to two VCPUs, among which the number of active ones depends on applications. The amount of memory allocated to a VM is set to 2 GB.

On the cloud testbed, we run the RUBiS [6] distributed online web service benchmark as an illustrating cloud service. RUBiS provides an auction site prototype modeling eBay.com and it is widely used as the benchmark program to evaluate the server performance and web application designs. The RUBiS servers form a three-tier server architecture consisted of the Web, application and database servers. RUBiS clients send requests with different workload patterns (browsing, bidding and mixed with adjustable composition of the two actions) to the Web server and simulate auctions of items on eBay.

To profile the application's resource demands in the cloud environment, we exploit third-party monitoring tools, sysstat [25] to collect runtime performance data in the hypervisor and VMs, and a modified perf [26] to obtain the values of performance counters from the Xen hypervisor on each server in the cloud testbed. In total, 518 metrics are profiled, i.e., 182 for the hypervisor and 182 for VMs by sysstat and 154 for performance counters by perf, periodically. They cover the statistics of every component of cloud servers, including the CPU usage, process creation, task switching activity, memory and swap space utilization, paging, interrupts, network activity, I/O and data transfer, power management, and more. Table 1 lists and describes a sampling of the performance metrics

Table 1. A sample of performance metrics used to characterize workload of the RUBiS benchmark system on the cloud testbed.

| Metric | Description |
|---|---|
| **%system** | Percentage of CPU utilization that occurred while executing at the system level. |
| **%user** | Percentage of CPU utilization that occurred while executing at the user level. |
| **%nice** | Percentage of CPU utilization that occurred while executing at the user level with nice priority. |
| **%iowait** | Percentage of time that the CPU or CPUs were idle during which the system had an outstanding disk I/O request. |
| **%soft** | Percentage of time spent by the CPU or CPUs to service software interrupts. |
| **%steal** | Percentage of time spent in involuntary wait by the virtual CPU or CPUs while the hypervisor was serving another virtual processor. |
| **proc/s** | Total number of tasks created per second. |
| **cswch/s** | Total number of context switches per second. |
| **intr/s** | Total number of interrupts received per second by the CPU. |
| **kbmemused** | Amount of used memory in kilobytes. |
| **kbbuffer** | Amount of memory used as buffers by the kernel in kilobytes. |
| **%memused** | Percentage of used memory. |
| **pswpin/s** | Total number of swap pages the system brought in per second. |
| **pswpout/s** | Total number of swap pages the system brought out per second. |
| **pgpgin/s** | Total number of kilobytes the system paged in from disk per second. |
| **pgpgout/s** | Total number of kilobytes the system paged out to disk per second. |
| **fault/s** | Number of page faults (major+minor) made by the system per second. |
| **pgsteal/s** | Number of pages the system has reclaimed from cache (pagecache and swapcache) per second to satisfy its memory demands. |
| **pgscank/s** | Number of page scanned by the kswapd daemon per second. |
| **%vmeff** | Calculated as pgsteal/pgscan, a metric of the efficiency of page reclaim. |
| **await** | The average time (in milliseconds) for I/O requests issued to the device to be served. |
| **tps** | Total number of transfer per second that were issued to physical devices. |
| **wr_sec/s** | Number of sectors written to the device. |
| **rd_sec/s** | Number of sectors read from the device. |
| **rxpck/s** | Total number of packets received per second. |
| **txpck/s** | Total number of packets transmitted per second. |
| **tcp-tw** | Number of TCP sockets in TIME$_W AIT state$. |
| **cycles** | Total number of CPU cycles. |
| **ITLB-load** | Total number of load operations to the instruction TLB. |
| **branches** | Number of branch operations. |
| **branch-misses** | Percentage of branch misses with the total number of branches. |
| **LLC-store-misses** | Number of last-level cache store misses operations. |
| **LLC-prefetches** | Number of last-level cache prefetches operations. |
| **Major-faults** | Number of page major faults. |
| **DTLB-load-misses** | Number of load misses operation to the data TLB. |
| **DTLB-stores** | Number of stores operation to the data TLB. |

that are used to characterize the workload dynamics of cloud applications on our testbed.

## 4. Experimental Results and Analysis

We run the RUBiS benchmark system on the cloud testbed and profile the workload dynamics with different clients' request patterns on both virtualized and non-virtualized environments. In this section, we present the results from the experiments and analyze them to find the workload characteristics and the impact of virtualization on the dynamics of resource demands.

### 4.1 Workload Characterization in a Virtualized Environment

In the first set of experiments, we deployed the RUBiS servers in VMs: the front-end Apache web server and PHP application server (The two servers are integrated together in the PHP implementation.) and the back-end MySQL database server. 1000 clients external to the cloud testbed sent browsing, bidding and mixed type requests to the web server. The think time was set to 7 second. We ran the experiments for around 20 minutes and profiled the resource demands for CPU, RAM, disk and network both in VMs and the hypervisor (*dom0*). Figures 1-4 depict the workloads. We tested five types of request compositions: browsing only, bidding only, 30% browsing and 70% bidding, 50% browsing and 50% bidding, and 70% browsing and 30% bidding. Due to the space limitation, we only include the results of the first two compositions in this paper.

The first two sub-figures in each set show the workload demands of the web and application servers and the database server for virtualized resources, including CPU cycles, amount of RAM, disk reads and writes, and data received and transmitted through networks in VMs. The last sub-figure in each set presents the overall workload demands to the physical resources.

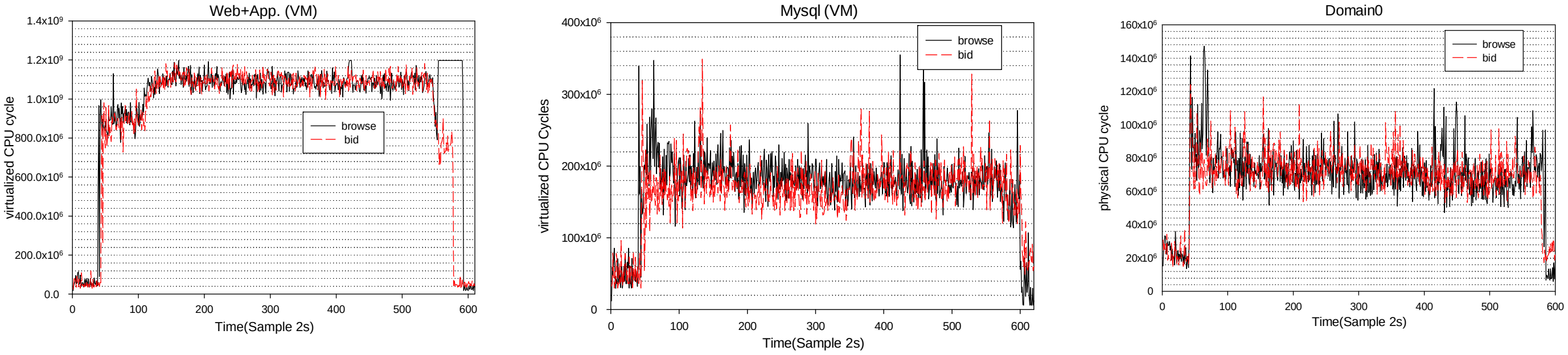

Figure 1. CPU cycle demands by the web and application servers and the database servers in VMs and the hypervisor (*dom 0*) to process the browsing and bidding requests from 1000 clients.

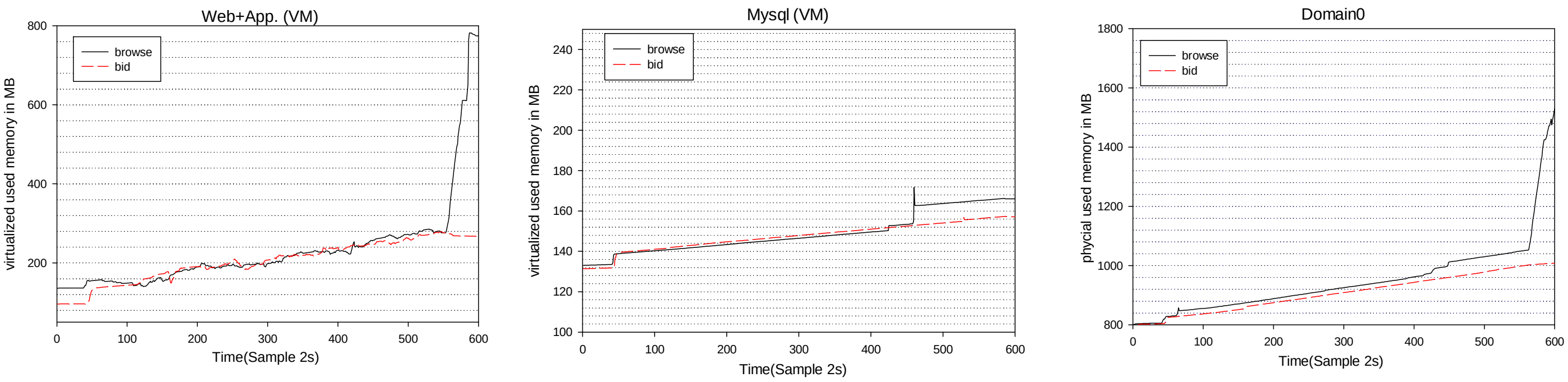

Figure 2. RAM demands by the web and application servers and the database servers in VMs and the hypervisor.

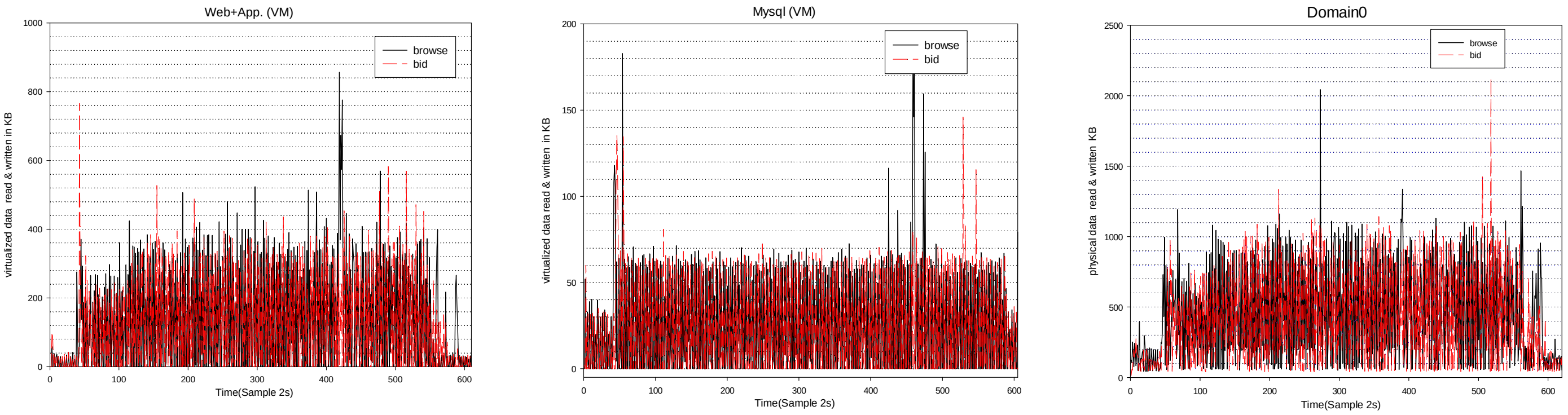

Figure 3. Disk read and write by the web and application servers and the database servers in VMs and the hypervisor.

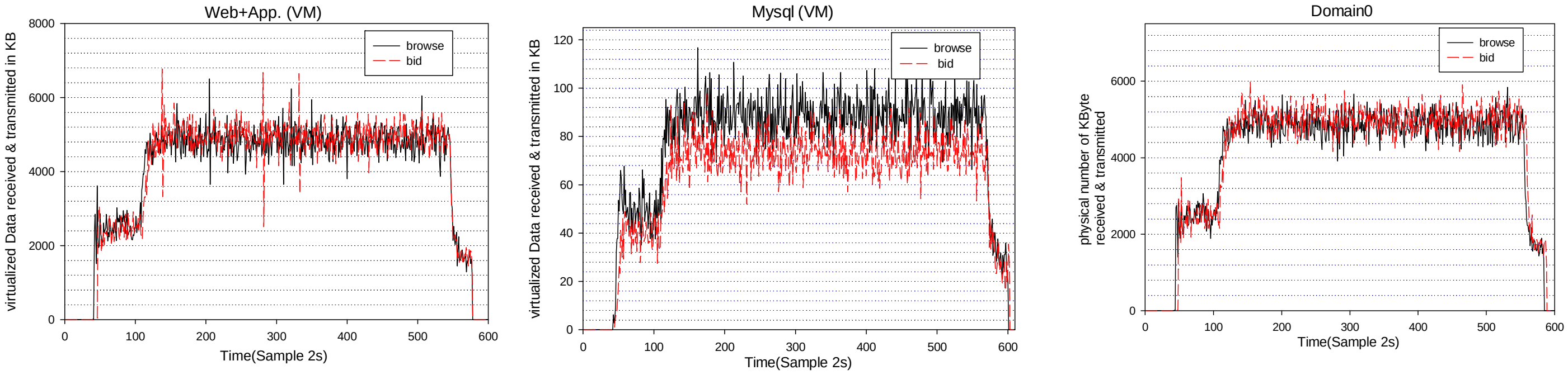

Figure 4. Network data received and transmitted by the web and application servers and the database servers in VMs and the hypervisor.

From the figures, we can see the workload curves for different types of resources display different shapes/ distributions with different means and variances. But for each type of resource, the workload dynamics show some patterns that can be quantified by formal models. In addition, there exist some lags between workload changes of the database server and the web and application servers as the client requests are received and processed first by the web server before being sent to the back-end database server. Between the front-end servers and back-end server, the front-end servers generate higher workload demands as they demand 6.11, 3.29, 5.71, and 55.56 times more CPU cycles, RAM space, disk read/write, and network data than the back-end server. When we compare the aggregated workload demands of the VMs with that of the hypervisor, the former is 16.84, 0.58, 0.47, and 0.98 times more/less than the latter with regard to the four types of resources. This indicates the hypervisor performs additional work other than the workload of RUBiS servers.

Comparing the two client request compositions, their workload dynamics display similar shapes except for the RAM demands. Figure 2 shows the browsing requests experience one or more jumps demanding more RAM, while the bidding requests have a more smooth curve. A possible explanation is that as more client browsing requests arrive, some requests are backlogged and after a certain period of time the server allocates more RAM to process those backlogged requests, which also causes more disk reads/writes (the spikes in the first two sub-figures of Figure 3). On the other hand, the longer think time of the bidding requests allows the servers to process the requests more smoothly. Another important finding is that although the browsing requests demand similar or more virtualized CPU and network resources than the bidding requests, the latter demands a little more physical resources than the former as shown in Figures 1 and 4.

### 4.2 Workload Characterization in a Non-Virtualized Environment

In order to characterize the impact of virtualization on system's workload, we conduct a series of experiment on non-virtualized servers in our testbed. The front-end web and application servers and the back-end database servers reside on separate physical servers. 1000 clients external to the RUBiS servers send browsing and bidding requests to the web server. *Sysstat* and *perf* profile resource usages directly from the host OS and hardware on each physical server. Figures 5-8 show the experimental results. The workload curves still display certain patterns that can be modeled.

We are interested in comparing the results with those from the virtualized environment as shown in Section 4.1. The two sets of figure show that the workload curves display the similar shapes and the front-end servers demand more resource than the back-end server. The aggregated demands for the four types of resources in the non-virtualized setting are 3.47, 0.97, 0.6 and 0.98 times more/less than those in the virtualized environment. The workload requests for RAM show the most significant difference between the two environments. As in the non-virtualized system (Figure 6), the bidding requests also display abrupt increase of RAM usage and the jumps happen earlier in time than those in the virtualized system. One reason is the longer communication delay in the non-virtualized system. In addition, from Figure 7 we can see disk read and write workload shows higher variance in the non-virtualized system than the virtualized one.

Comparing the results in Sections 4.1 and 4.2, we find application's demand for physical resources is higher in the non-virtualized environment than in the virtualized one, with 88% more CPU cycles, 21% more RAM, and 2% more network traffic, while disk read/write is 25% less. These findings will allow cloud service providers to achieve efficient capacity planning for a desirable SLA satisfaction rate.

## 5. Conclusion

It is imperative to understand the application/ service workload characteristics in the cloud for effective resource planning and management. In this work, we study the impact of virtualization on the workload dynamics. We present experimental results on a cloud testbed by profiling the workload dynamics on both virtualized and non-virtualized environments. We compare the resource demands at the three server tiers.

This study is preliminary. Our goal is to extract the rules of thumb to aid cloud service providers to achieve the best resource planning. We plan to design and apply formal methods to model the workload dynamics at both resource level and transaction level. We also plan to characterize the workload of other cloud applications, such as big data applications using the MapReduce paradigm.

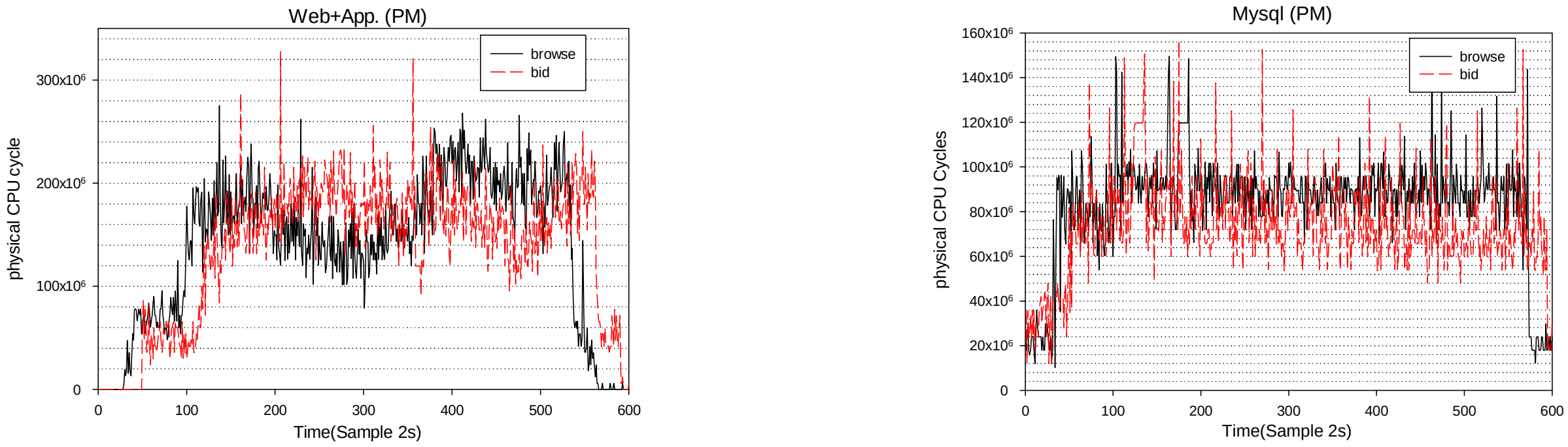

Figure 5. CPU cycle demands by the web and application servers and the database servers to process the browsing and bidding requests from 1000 clients.

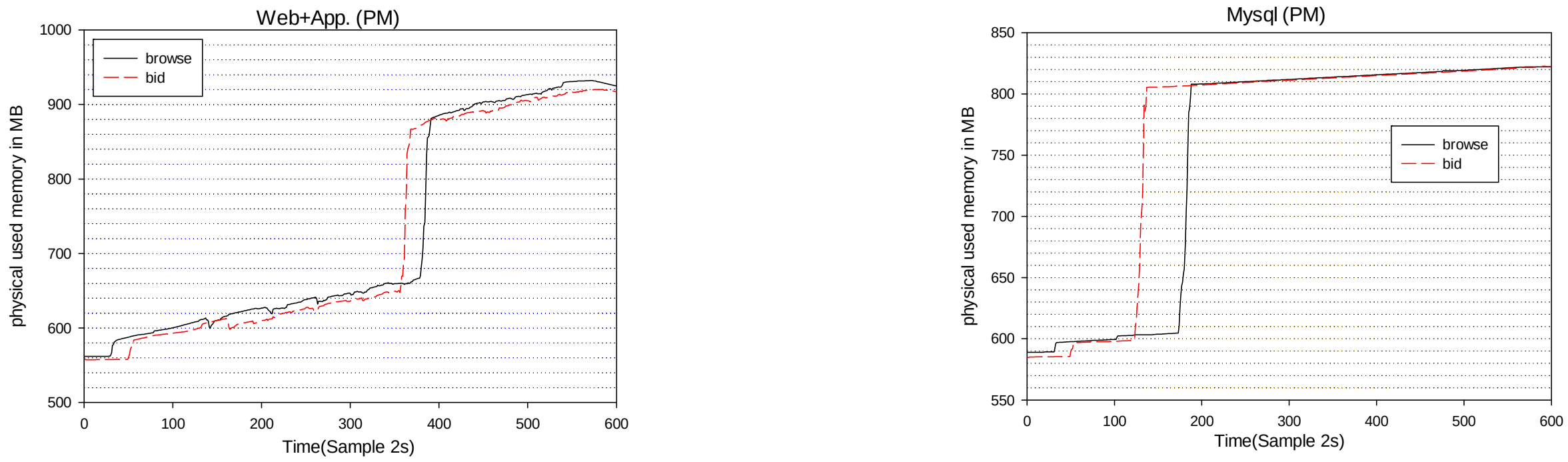

Figure 6. RAM demands by the web and application servers and the database servers.

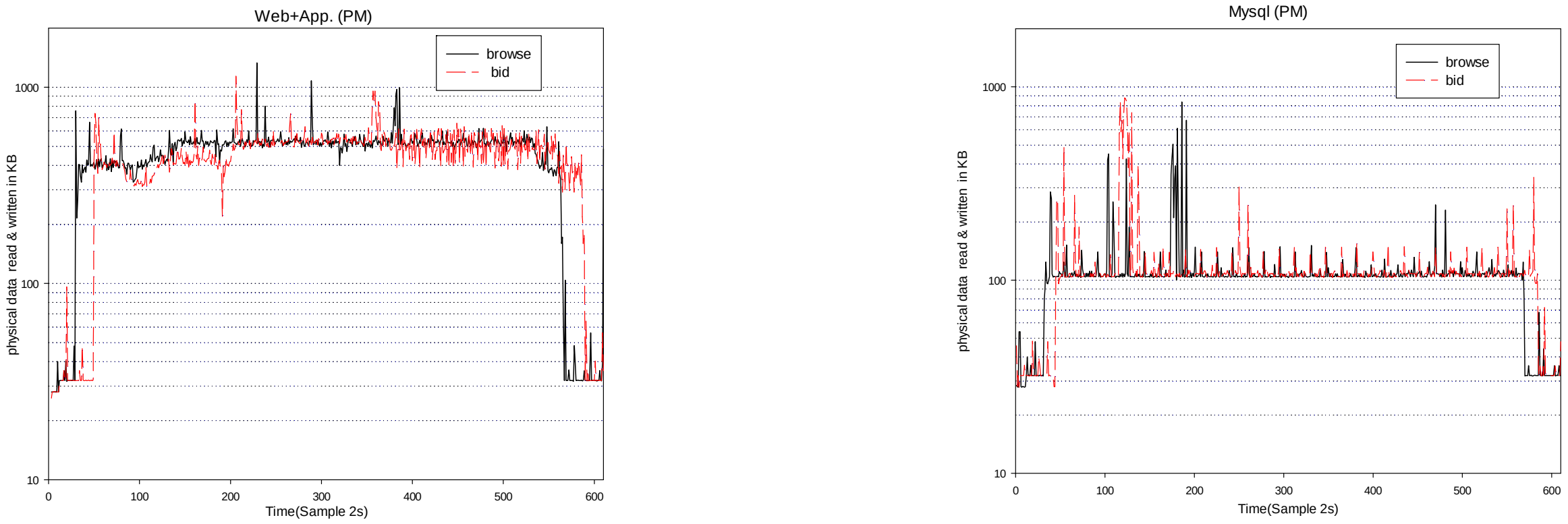

Figure 7. Disk read and write by the web and application servers and the database servers.

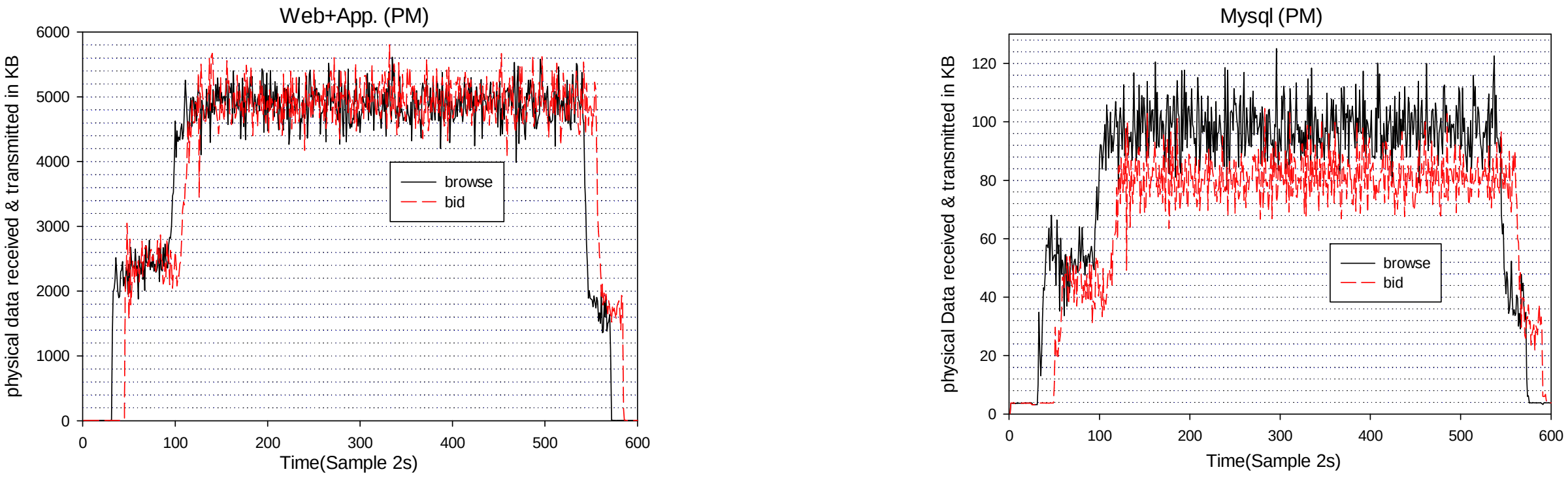

Figure 8. Network data received and transmitted by the web and application servers and the database servers.

## Acknowledgment

We would like to thank the anonymous reviewers for their constructive comments and suggestions. This work was performed in the Dependable Computing Systems Laboratory at the University of North Texas.

## References

[1] M. Armbrust, A. Fox, R. Griffith, A. D. Joseph, R. Katz, A. Konwinski, G. Lee, D. Patterson, A. Rabkin, I. Stoica, and M. Zaharia. A view of cloud computing. *Communications of the ACM*, 53(4):50–58, 2010.

[2] J. E. Smith and R. Nair. The architecture of virtual machines. *IEEE Computer*, 38(5):32–38, 2005.

[3] Z. Wang, X. Tang, X. Luo. Policy-Based SLA-Aware Cloud Service Provision Framework. In *Proceedings* of *IEEE International Conference on Semantics Knowledge and Grid (SKG)*, 2011.

[4] Q. Guan, C. Chiu, S. Fu. CDA: A Cloud Dependability Analysis Framework for Characterizing System Dependability in Cloud Computing Infrastructures. In *Proceedings of IEEE the 18th International Symposium on Dependable Computing (PRDC)*, 2012.

[5] G. Wang and T. S. Eugene Ng. The Impact of Virtualization on Network Performance of Amazon EC2 Data Center. In *Proceedings of IEEE Conference on Computer Communications (INFOCOM)*, 2010.

[6] RUBiS Website. http://rubis.ow2.org

[7] E. Hernández-Orallo and J. Vila-Carb. "Web server performance analysis using histogram workload models," *Computer Networks*, vol. 53, no. 15, pp. 2727–2739, 2009.

[8] W. Shi, Y. Wright, E. Collins, and V. Karamcheti. "Workload characterization of a personalized web site and its implications for dynamic content caching," In *Proceedings of International Conference on Web Content Caching and Distribution (WCW)*, 2002.

[9] E. Thereska, A. Donnelly, and D. Narayanan, "Sierra: practical powerproportionality for data center storage," In *Proceedings of ACM European Conference on Computer Systems (EuroSys)*, 2011.

[10] L. N. Bairavasundaram, A. C. Arpaci-Dusseau, R. H. Arpaci-Dusseau, G. R. Goodson, and B. Schroeder, "An analysis of data corruption in the storage stack,", *ACM Transactions on Storage*, vol. 4, no. 3, pp. 821–834, 2008.

[11] F. Wang, Q. Xin, B. Hong, S. A. Brandt, E. L. Miller, D. D. E. Long, and T. T. Mclarty, "File system workload analysis for large scale scientific computing applications," in *Proceedings of IEEE Conference on Mass Storage Systems and Technologies (MSST)*, 2004.

[12] D. Ersoz, M. S. Yousif, and C. R. Das, "Characterizing network traffic in a cluster-based, multi-tier data center," in *Proceedings of IEEE International Conference on Distributed Computing Systems (ICDCS)*, 2007.

[13] V. Paxson, "Empirically derived analytic models of wide-area TCP connections", *IEEE/ACM Transactions of Networking*, vol. 2, no. 4, pp. 316–336, 1994.

[14] K. Christodoulopoulos, V. Gkamas, and E. A. Varvarigos, "Statistical analysis and modeling of jobs in a grid environment," *Journal of Grid Computing* , vol. 6, no. 1, 2008.

[15] E. Medernach, "Workload analysis of a cluster in a grid environment," in *Proceedings of Job Scheduling Strategies for Parallel Processing Workshop in conjunction with IEEE International Parallel and Distributed Processing Symposium (IPDPS)*, 2005.

[16] B. Song, C. Ernemann, and R. Yahyapour, "User group-based workload analysis and modelling," in *Proceedings of IEEE/ACM International Symposium on Cluster, Cloud and Grid Computing (CCGrid)*, 2005.

[17] A. Iamnitchi, S. Doraimani, and G. Garzoglio, "Workload characterization in a high-energy data grid and impact on resource management," *Cluster Computing*, vol. 12, no. 2, pp. 153–173, 2009.

[18] L. Wang, J. Zhan, C. Luo, Y. Zhu, Q. Yang, Y. He, W. Gao, Z. Jia, Y. Shi, S. Zhang, C. Zheng, G. Lu, K. Zhan, X. Li, and B. Qiu, "BigDataBench: a big data benchmark suite from Internet services," in *Proceedings of IEEE International Symposium on High Performance Computer Architecture (HPCA)*, 2014.

[19] C. Luo, J. Zhan, Z. Jia, L. Wang, G. Lu, L. Zhang, C. Xu, and N. Sun, "CloudRank-D: benchmarking and ranking cloud computing systems

for data processing applications,”, *Frontiers of Computer Science*, vol. 6, no. 4, pp. 347-362, 2012.

[20] A. D'Ambrogio, P. Bocciarelli. “A model-driven approach to describe and predict the performance of composite services”. In *Proceedings of ACM International Workshop on Software and Performance (WOSP)*, 2007.

[21] S. Kavulya, J. Tan, R. Gandhi, and P. Narasimhan, “An analysis of traces from a production mapreduce clustePr,” in *Proceedings of IEEE/ACM International Symposium on Cluster, Cloud and Grid Computing (CCGrid)*, 2010.

[22] A. K. Mishra, J. L. Hellerstein, W. Cirne, and C. R. Das, “Towards characterizing cloud backend workloads: insights from google compute clusters,” *SIGMETRICS Performance Evaluation Review*, vol. 37, no. 4, pp. 34–41, 2010.

[23] B. F. Cooper, A. Silberstein, E. Tam, R. Ramakrishnan, and R. Sears, “Benchmarking cloud serving systems with YCSB,” In *Proceedings of ACM Symposium on Cloud Computing (SOCC)*, 2010.

[24] L. Wang, J. Tao, M. Kunze, A. C. Castellanos, D. Kramer, W. Karl. Scientific Cloud Computing: Early Definition and Experience. In *Proceedings of IEEE International Conference on High Performance Computing and Communications (HPCC)*, 2008.

[25] Sysstat. http://sebastien.godard.pagesperso-orange.fr/.

[26] Perf. http://en.wikipedia.org/wiki/ Perf_(Linux)/.